\documentclass[preprint,authoryear]{elsarticle}
\usepackage{amssymb,natbib}
\journal{New Astronomy}
\def\astrobj#1{#1}

\begin{document}

\begin{frontmatter}

\title{PPMXL photometric analysis of three open cluster candidates}

\author{A. L. Tadross $^*$}

\address{National Research Institute of Astronomy and Geophysics, Helwan, Cairo, Egypt,\\ email: altadross@yahoo.com}
\cortext[cor*]{Tel/Fax: +202 25560645/ +202 25548020}

\begin{abstract}
 We present here the astrophysical parameters of three stellar open cluster candidates using PPMXL\footnote{\it http://vizier.cfa.harvard.edu/viz-bin/VizieR?-source=I/317} database. In this study, the main photometric, astrometry and statistical parameters of \astrobj{Ruprecht 13}, \astrobj{Ruprecht 16} and \astrobj{Ruprecht 24} are estimated for the first time.
\end{abstract}

\end{frontmatter}

Open star clusters (OCs) play an important role in studying the formation and evolution of the Galactic disk and the stellar evolution as well. The fundamental physical parameters of OCs, e.g. distance, reddening, age, and metallicity are necessary for studying the Galactic disk. The strong interest of OCs results come from their fundamental properties. Among the 1787 currently OCs, more than half of them have been poorly studied or even unstudied up to now, Piatti et al. (2011). Thus the current paper is a part of our continuation series whose goal is to obtain the main astrophysical properties of previously unstudied OCs using modern databases. The most important thing for using PPMXL database lies in containing the positions, proper motions of USNO-B1.0\footnote{\it http://vizier.cfa.harvard.edu/viz-bin/VizieR?-source=I/284} and the Near Infrared (NIR) photometry of the Two Micron All Sky Survey (2MASS)\footnote{\it http://vizier.cfa.harvard.edu/viz-bin/VizieR?-source=II/246}, which let it be the powerful detection of the star clusters behind the hydrogen thick clouds on the Galactic plane.

Our candidates are selected from among the unstudied OCs of Ruprecht catalogue (hereafter, \astrobj{Ru 13}, \astrobj{Ru 16}, and \astrobj{Ru 24}). They are lying in the third quarter of Galactic longitude and close to the Galactic plane. The only available information about these clusters are the coordinates and the apparent diameters, which were obtained from WEBDA\footnote{\it http://obswww.unige.ch/webda} site and the last updated version of DIAS\footnote{\it http://www.astro.iag.usp.br/$\sim$wilton/} collection (version 3.0, 2010 April 30). These clusters are sorted by right ascensions and listed in Table 1. The quality of the data is taken into account and the physical properties of each cluster are estimated by applying the same methodology.

This paper is organized as follows. PPMXL data extraction and preparation are presented in Section 2, while the data analysis and parameters estimations are described in Sections 3. Finally, the conclusion is devoted to Section 4.

\begin{table}
\caption{Equatorial, Galactic positions and  estimated diameters of the clusters under investigation; sorted by right ascensions.}
\tabcolsep 5.5 pt
\centerline{
\begin{tabular}{ccrrrrr}\\
\hline
            Cluster
            & $\alpha~^{h}~^{m}~^{s}$
            & $\delta~^{\circ}~{'}~{''}$
            & G. Long.$^{\circ}$
            & G. Lat.$^{\circ}$
            & $D.^{'}$\\
\hline
\astrobj{Ru 13} &  07:07:48  &  --25:52:01 & 237.88  & --8.153 & 5.0\\
\astrobj{Ru 16} &  07:23:12  &  --19:27:00 & 233.79  & --2.058 & 4.0\\
\astrobj{Ru 24} &  07:31:54  &  --12:45:00 & 228.89  & 2.971 & 7.0\\
\hline\\
\end{tabular}}
\end{table}
%---------------------------------------------------------------------
\section*{2. PPMXL Data Extraction and preparation}

The astrophysical parameters of the investigated cluster have been estimated using the PPMXL Catalogue of Roeser et al. (2010). It is combining the USNO-B1.0 proper motion of Monet et al. (2003) and NIR JH$K_{s}$ pass-band of 2MASS Point Sources of Cutri et al. (2003). USNO-B1.0 is a very useful catalogue, which gives us an opportunity to distinguish between the members and background/field stars. While, NIR 2MASS photometry provides J (1.25$\mu$m), H (1.65$\mu$m), and $K_{s}$ (2.17$\mu$m) band photometry for millions of galaxies and nearly a half-billion stars (Carpenter, 2001). This survey has proven to be a powerful tool in the analysis of the structure and stellar content of open clusters (Bica et al. 2003, Bonatto \& Bica 2003). The photometric uncertainty of the 2MASS data is less than 0.155 at $K_{s} \sim 16.5$ magnitude which is the photometric completeness for stars with $|b| > 25^{o}$, Skrutskie et al. (2006).

It is noted that the candidate clusters are located near the Galactic plane ($|b|<10~^{o}$), therefore we expect significant foreground and background field contamination. These clusters  have low central concentration, as appeared from their images on the Digitized Sky Survey (DSS)\footnote{\it http://cadcwww.dao.nrc.ca/cadcbin/getdss}, see Fig. 1. Their apparent diameters are less than 10 arcmin, hence the downloaded data have been extended to reach the field background stars, whereas the clusters dissolved there, i.e. the data are extracted at such sizes of about 20 arcmin.

\begin{figure*}
\begin{center}
{\includegraphics[width=12cm]{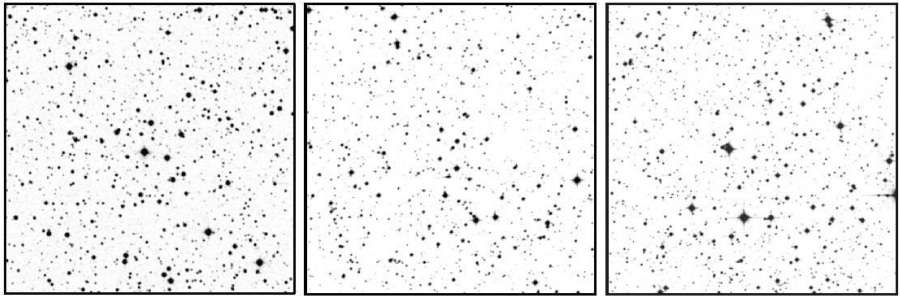}}
\end{center}
\caption{The blue images of the candidate clusters. From left to right: \astrobj{Ru 13}, \astrobj{Ru 16} \& \astrobj{Ru 24} respectively, as taken from Digitized Sky Surveys. North is up, east on the left.}
\end{figure*}

Comparing the data with Cutri et al.'s Point Source Catalogue (2003), we found that most stars' magnitudes have a ``AAA'' quality flag, which means the Signal Noise Ratio is $SNR \geq 10$, i.e. they have the highest quality measurements. Every star in our data has 3-colour photometric values $J, H, K_{s}$ mag; and proper motion values in right ascension and declination, i.e. (pm $\alpha$ cos $\delta$) and (pm $\delta$) mas yr$^{-1}$. According to Roeser et al. (2010), the stars with proper motion uncertainties $\geq$ 4.0 mas yr$^{-1}$ have been removed. Also, in this context, the stars with observational uncertainties $\geq$ 0.20 mag are excluded, and the photometric completeness limit is applied on the photometric pass-band 2MASS data to avoid the over-sampling at the lower parts of the cluster's colour magnitude diagrams (CMDs) (cf. Bonatto et al. 2004). Pm vector point diagram (VPD) with distribution histogram of 5 mas yr$^{-1}$ bins have been constructed as shown in Fig. 2. The Gaussian function fit to the central bins provides the mean pm in both directions. All data lie at that mean $\pm 1~\sigma$ (where $\sigma$ is the standard deviation of the mean) can be considered as astrometric probable members. In addition, the photometric membership criteria is adopted based on the location of the stars within $\pm 0.1$ mag around the ZAMS curve in the CMDs (Claria \& Lapasset 1986).

\begin{figure*}
\begin{center}
{\includegraphics[width=12cm]{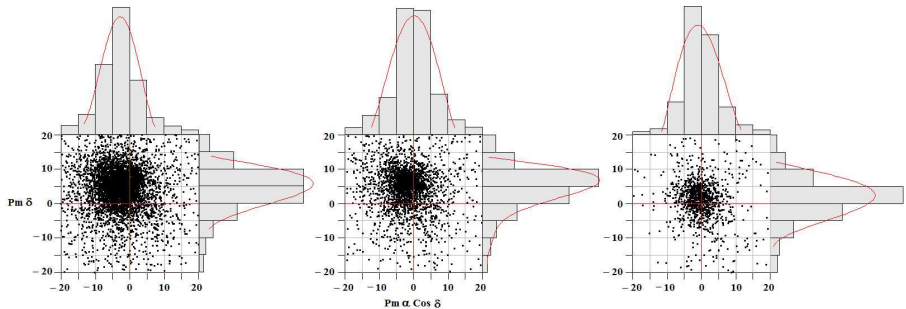}}
\end{center}
\caption{Proper motion vector point diagrams {\it VPDs}~ for our candidate clusters after avoiding all data with pm errors $\geq$ 4 mas yr$^{-1}$. From left to right: \astrobj{Ru 13}, \astrobj{Ru 16} \& \astrobj{Ru 24} respectively. Histograms of 5 mas yr$^{-1}$ bins in both directions are drawn. The Gaussian function fit to the central bins provides the mean values of pm $\alpha$ cos $\delta$ = (--2.6 $\pm$ 0.1), (--2.0 $\pm$ 0.09) \& (--0.6 $\pm$ 0.06) mas yr$^{-1}$ and pm $\delta$ = (4.9 $\pm$ 0.05), (5.5 $\pm$ 0.1) \& (1.6 $\pm$ 0.06) mas yr$^{-1}$ respectively.}
\end{figure*}
%----------------------------------------------------------------
\section*{3. Data Analysis}
\subsection*{3.1. Cluster's Centre and Radial Density Profile}

The centre of any cluster can be roughly estimated by eye, but to determine the centre's coordinates of our candidates more precisely, we applied the star-count method to the whole area of each cluster. All the data that obtained around the adopted centre are dividing into equal sized bins in $\alpha$ and $\delta$. The cluster centre is define as the location of maximum stellar density of the cluster's area. The Gaussian curve-fitting is applied to the profiles of star counts in $\alpha$ \& $\delta$ respectively (cf. Tadross 2011). The estimated clusters' centres are found to be in agreement with that obtained from Webda with 3$^{s}$ in $\alpha$ and 10$^{''}$ in $\delta$.

To establish the radial density profile (RDP) of a cluster, we counted the stars within concentric shells in equal incremental steps of $r\leq1$ arcmin from the cluster centre. We repeated this process for $1<r\leq2$ up to $r\leq20$ arcmin, i.e. the stellar density is derived out to the preliminary radius. The stars of the next steps should be subtracted from the later ones, so that we obtained only the amount of the stars within the relevant shell's area, not a cumulative count. Finally, we divided the star counts in each shell to the area of that shell those stars belong to. The density uncertainties in each shell was calculated using Poisson noise statistics.

Fig. 3 shows the RDP from the centre of the cluster to a maximum angular separation of 10 arcmin for the investigated clusters. To determine the structural parameters of the cluster more precisely, we applied the empirical King model (1966). The King model parameterizes the density function $\rho(r)$ as:
\begin{equation}
\rho(r)=f_{bg}+\frac{f_{0}}{1+(r/r_{c})^{2}}
\end{equation}

where $f_{bg}$, $f_{0}$ and $r_{c}$ are background, central star density and the core radius of the cluster, respectively. From the concentration parameter $c$, defined as $c= (R_{lim}/R_{core})$, Nilakshi et al. (2002) concluded that the angular size of the coronal region is about 6 times the core radius. Maciejewski \& Niedzielski (2007) reported that $R_{lim}$ may vary for individual clusters between about $2 R_{core}$ and $7 R_{core}$. In our case, from the $c$
values, we can see that $R_{lim}$ vary between $5.8 R_{core}$ and $6.7 R_{core}$, i.e. tending to the upper limit of Maciejewski \& Niedzielski (2007).
The cluster limited radius can be defined at that radius which covers the entire cluster area and reaches enough stability with the background density, i.e. the difference between the observed density profile and the background one is almost equal zero. It is noted that the determination of a cluster radius is made by the spatial coverage and uniformity of PPMXL photometry which allows one to obtain reliable data on the projected distribution of stars for large extensions to the clusters' halos. On the other hand, the concentration parameter seems to be related to cluster age, i.e. for clusters younger than about 1 Gyr, it tends to increase with cluster age. Nilakshi et al. (2002) notes that the halos' sizes are smaller for older systems. Finally, we can say that open clusters appear to be larger in the near-infrared than in the optical data, Sharma et al. (2006).\\
Knowing the cluster's total mass (Sec. 3.3), the tidal radius can be given by applying the equation of Jeffries et al. (2001):
\begin{equation}
R_{t} = 1.46 ~ (M_{c})^{1/3}
\end{equation}
where $R_{t}$ and $M_{c}$ are the tidal radius and total mass of the cluster respectively.
%----------------------------------------------------------------

\begin{figure*}
\begin{center}
{\includegraphics[width=12cm]{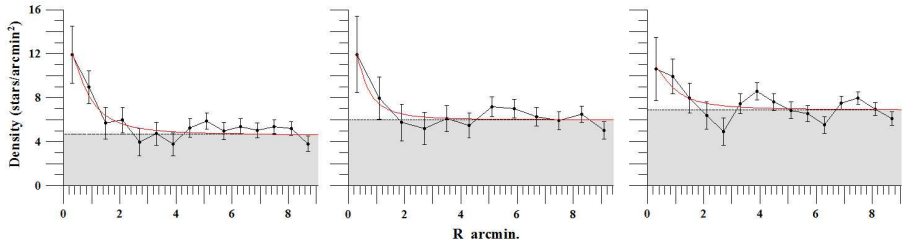}}
\end{center}
\caption{The radial density distribution for stars in the field of the candidate clusters. From left to right:\astrobj{Ru 13}, \astrobj{Ru 16} \& \astrobj{Ru 24} respectively. The density shows maximum value at the cluster' centre and then decreases down to that point at which the decrease becomes asymptotically. The curved solid lines represent the fitting of King (1966) model. Errors bars are determined from sampling statistics ($1/\sqrt{N}$, where N is the number of stars used in the density estimation at each point). The limited radii $R_{lim}$ = 4.5, 3.5 \& 5.0 arcmin; and the core radii $r_{c}$ = 0.77, 0.52 \& 0.83 arcmin respectively. The dark areas refer to the background field densities where $f_{bg}$ = 4.6, 6.0 \& 6.8 stars per arcmin$^{2}$ respectively.}
\end{figure*}

%----------------------------------------------------------------
\subsection*{3.2. Colour-Magnitude Diagrams}

The main photometrical parameters (age, distance, and reddening) are determined for our clusters by fitting the 2MASS ZAMS solar metallicity isochrones of Marigo et al. (2008) to both CMDs (J, J-H \& K$_{s}$, J-K$_{s}$) of each cluster. Marigo et al. (2008) isochrones can be selected and downloaded from the site of Padova isochrones\footnote{http://stev.oapd.inaf.it/cgi-bin/cmd}. Several isochrones of different ages are applied to each cluster; the best fit should be obtained at the same distance modulus for both diagrams, and the colour excesses should be obeyed Fiorucci \& Munari (2003)'s relations for normal interstellar medium, see Fig. 4.

The observed data has been corrected for interstellar reddening using the coefficients ratios $\frac {A_{J}}{A_{V}}=0.276$ and $\frac {A_{H}}{A_{V}}=0.176$, which were derived from absorption rations of Schlegel et al. (1998), while the ratio $\frac {A_{K_s}}{A_{V}}=0.118$ was derived from Dutra et al. (2002).

Fiorucci \& Munari (2003) calculated the colour excess values for 2MASS photometric system. We ended up with the following results: $\frac {E_{J-H}}{E_{B-V}}=0.309\pm0.130$, $\frac {E_{J-K_s}}{E_{B-V}}=0.485\pm0.150$, where R$_{V}=\frac {A_{V}}{E_{B-V}}= 3.1$. Also, we can de-reddened the distance moduli using these formulae:  $\frac {A_{J}}{E_{B-V}}$= 0.887, $\frac {A_{K_s}}{E_{B-V}}$= 0.322, then the distance of each cluster from the Sun ($R_{\odot}$) can be calculated.

Once the cluster's distance $R_{\odot}$ is estimated, then the distance from the galactic centre ($R_{G}$) and the projected distances on the galactic plane from the Sun ($X_{\odot}~\&~Y_{\odot}$) and the distance from galactic plane ($Z_{\odot}$) can be determined, (for more details about the distances calculations, see Tadross 2011).

\begin{figure*}
\begin{center}
{\includegraphics[width=12.1cm]{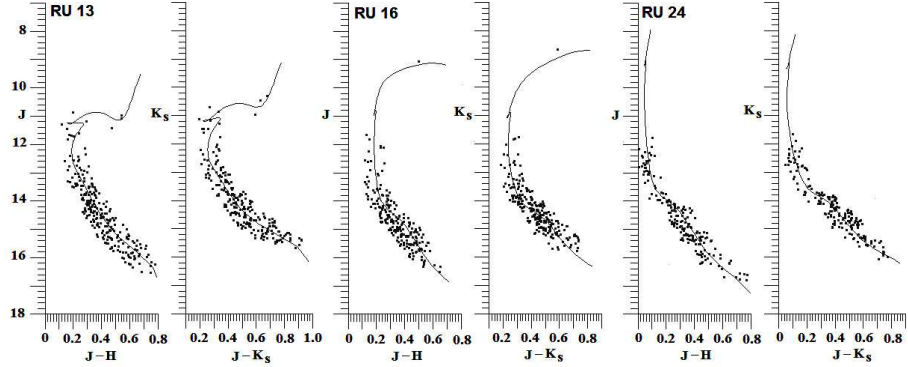}}
\end{center}
\caption{The net NIR CMDs of \astrobj{Ru 13}, \astrobj{Ru 16} \& \astrobj{Ru 24} for stars lying closely to the fitted isochrones and after removing all contaminated field stars. Age, distance modulus, E(J-H) \& E(J-K$_{s}$) are estimated for each cluster as (1.0 Gyr, 10.8, 0.08 \& 0.12 mag), (160 Myr, 12.3, 0.22 \& 0.31 mag) and (60 Myr, 11.8, 0.11 \& 0.18 mag) respectively.}
\end{figure*}
%----------------------------------------------------------------
\subsection*{3.3. Luminosity, Mass Functions and the total mass}

Since the central concentration of the cluster is relatively low, the determination of the membership using the stellar RDP is difficult. Even though we determine the membership within a circle of radius $r=2 \sim 3~r_{c}$ arcmin. By doing that, we obtained a more precise main-sequence in the CMDs. All of these stars are found very close to the main-sequence (MS) curve merged with some contaminated stars. These MS stars are very important in determining the luminosity and mass functions.

The luminosity function (LF) gives the number of stars per luminosity interval, or in other words, the number of stars in each magnitude bin of the cluster. It is used to study the properties of large groups or classes of objects, such as the stars in clusters or the galaxies in the Local Group.
In order to estimate the LF, we count the observed stars in terms of absolute magnitude after applying the distance modulus as shown in Fig. 5. The magnitude bin intervals are selected to include a reasonable number of stars in each bin and for the best possible statistics of the luminosity and mass functions. From LF, we can infer that the massive bright stars seem to be centrally concentrated more than the low masses and fainter ones (Montgomery et al. 1993).

\begin{figure*}
\begin{center}
{\includegraphics[width=11cm]{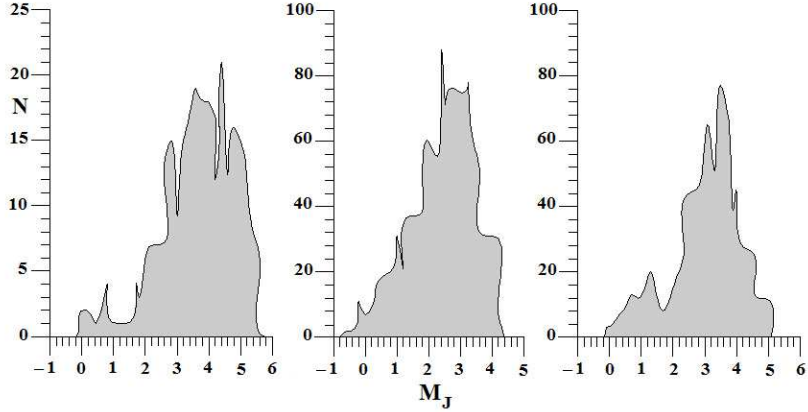}}
\end{center}
\caption{Spatial distribution of luminosity function for the clusters under investigation in terms of the absolute magnitude $M_{J}$. From left to right: \astrobj{Ru 13}, \astrobj{Ru 16} and \astrobj{Ru 24} respectively. The colour and magnitude filters cutoffs have been applied to each cluster. The peak values lie at $M_{J}$ = 4.4, 2.0 \& 3.5 mag; and the total luminosity are found to be -1.9, -4.5 \& 3.8 mag respectively.}
\end{figure*}

As known the LF and mass function (MF) are correlated to each other according to the Mass-luminosity relation. The accurate determination of both of them (LF \& MF) suffers from some problems e.g. the field contamination of cluster members; the observed incompleteness at low-luminosity (or low-mass) stars; and mass segregation, which may affect even poorly populated, relatively young clusters (Scalo 1998). On the other hand, while, the physical properties and evolutions of the stars are related to their masses, so the determination of the initial mass function (IMF) is needed, that is an important diagnostic tool for studying large quantities of star clusters. IMF is an empirical relation that describes the mass distribution (a histogram of stellar masses) of a population of stars in terms of their theoretical initial mass (the mass they were formed with). The IMF is defined in terms of a power law as following:
\begin{equation}
\frac{dN}{dM} \propto M^{-\alpha}
\end{equation}

where $\frac{dN}{dM}$ is the number of stars of mass interval (M:M+dM) within a specified volume of space, and $\alpha$ is a dimensionless exponent. The IMF for massive stars ($>$ 1 $M_{\odot}$) has been studied and well established by Salpeter (1955), where $\alpha$ = 2.35. This form of Salpeter shows that the number of stars in each mass range decreases rapidly with increasing mass. It is noted that our investigated clusters have MF slopes ranging around Salpeter's value, as shown in Fig. 6.

The mass of each star in the investigated clusters has been estimated from a polynomial equation developed from the data of the solar metallicity isochrones (absolute magnitudes vs. actual masses) at a specific age of each cluster individually. The summation of multiplying the number of stars in each bin by the mean mass of that bin yields the total mass of the cluster. Therefore, our clusters are found to have total masses of 230, 410 \& 240 {\it M$_{\odot}$} for \astrobj{Ru 13}, \astrobj{Ru 16} \& \astrobj{Ru 24} respectively.

\begin{figure*}
\begin{center}
{\includegraphics[width=11.5cm]{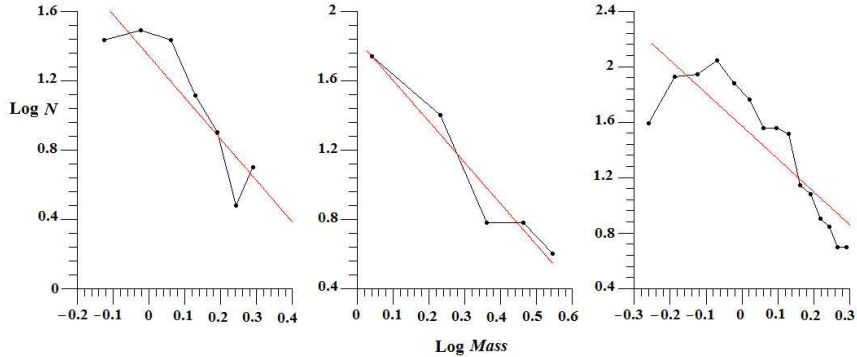}}
\end{center}
\caption{The mass function of the candidate clusters. From left to right: \astrobj{Ru 13}, \astrobj{Ru 16} \& \astrobj{Ru 24} respectively. The slope of the initial mass function {\it IMF}~ is found to be $\Gamma$ = -2.39, -2.37 \& -2.37 respectively.}
\end{figure*}
%---------------------------------------------------------------------

\subsection*{3.4. Dynamical state}

The relaxation time ($T_{relax}$) of a cluster is defined as the time in which the cluster needs from the very beginning to build itself and reach the stability state against the contraction and destruction forces, e.g. gas pressure, turbulence, rotation, and the magnetic field (cf. Tadross 2005).  $T_{relax}$ is depending mainly on the number of members and the cluster diameter. To describe the dynamical state of a cluster, the relaxation time can be calculated in the form:
\begin{equation}
 T_{relax}=\frac{N}{8\ln N}~T_{cross}\, , \;
\end{equation}
where $T_{cross}=D/\sigma_{V}$ denotes the crossing time, $N$ is the total number of stars in the investigated region of diameter $D$, and $\sigma_{V}$ is the velocity dispersion (Binney \& Tremaine  1998) with a typical value of 3 km s$^{-1}$ (Binney \& Merrifield 1987). Using the above formula we can estimate the dynamical relaxation time, and then the dynamical-evolution parameter $\tau$ can be calculate for each cluster by:
\begin{equation}
 \tau=\frac{age}{t_{relax}}\, , \;
\end{equation}
If the cluster's age is found greater than its relaxation time, i.e.  $\tau \gg$ 1.0, then the cluster was dynamically relaxed, and vice versa. In our case, all our clusters are found dynamically relaxed, where $\tau$ = 5.8, 6.7 \& 6.0 for \astrobj{Ru 13}, \astrobj{Ru 16} \& \astrobj{Ru 24} respectively.

%---------------------------------------------------------------------------
\section*{4. Conclusions}

Our procedure analysis has been applied for estimating the astrophysical parameters of three yet unstudied open clusters (\astrobj{Ru 13}, \astrobj{Ru 16} \& \astrobj{Ru 24}). Hence, we found that these clusters have reasonable stellar density profiles, lying in the same absolute distance modulus, reddening range, and having IMF slopes around the Salpeter's (1955) value. On the other hand, the ages of these clusters are found to be greater than their relaxation times that infer that these clusters are indeed dynamically relaxed. All parameters of the investigated clusters are listed in Table 2.
%---------------------------------------------------------------------
\section*{Acknowledgements}

It is worthy to mention that, this publication made use of WEBDA, DIAS catalogues, and the data products from PPMXL database of Roeser et al. (2010).
%---------------------------------------------------------------------

%---------------------------------------------------------------------
\begin{table*}
\caption{The derived astrophysical parameters for the investigated clusters. Columns display, respectively, cluster name, equatorial coordinates, age, distance modulus, distance from the sun, reddening values, limiting radius, core radius, tidal radius, distance from the Galactic centre, projected distances on the Galactic plane from the sun, X \& Y, distance from Galactic plane, total mass, number of members, time of relaxation, and background field density.}
\tabcolsep 3 pt
\centerline{
\begin{tabular}{ccccrrrrrrrrrrrrrrr}\\
\hline
            Cluster ~~~~& $\alpha$ & ~~~$\delta$ & ~~~Age  & ~~~m-M  & $R_{\odot}$ & ~~~~E$_{B-V}$ \\
             ~~~~& $^{h}~^{m}~^{s}$ & ~~~$^{\circ}~{'}~{''}$ & ~~~\it Myr & ~~~\it mag & \it pc & ~~~~\it mag \\
\hline
\astrobj{Ru 13} ~~~~& 07:07:51 & ~~~--25:52:11 & 1000 & ~~~10.8 & ~~~1300$\pm60$ & ~~~~0.26$\pm0.05$ \\
\astrobj{Ru 16} ~~~~& 07:23:13 & ~~~--19:27:05 & 160  & ~~~12.3 & ~~~2160$\pm100$ & ~~~~0.71$\pm0.07$ \\
\astrobj{Ru 24} ~~~~& 07:31:53 & ~~~--12:45:49 & 60   & ~~~11.8 & ~~~1983$\pm90$ & ~~~~0.35$\pm0.05$ \\
\hline\\
\end{tabular}}
\end{table*}
\begin{table*}
$Continued$\\
\tabcolsep 3 pt
\centerline{
\begin{tabular}{ccccrrrrrrrrrrrrrrr}\\
\hline
           $R_{lim}$ & ~R$_{core}$ &  ~R$_{tid}$ &  ~$R_{g}$ & ~$X_{\odot}$  & ~$Y_{\odot}$  & ~$Z_{\odot}$ & ~Mass & ~Mem. & ~$T_{relax}$ & ~$f_{bg}$ \\
           \it $^{\prime}$  & ~\it $^{\prime}$  & ~\it pc & ~\it kpc  & ~\it pc  & ~\it pc & ~\it pc & ~\it M$_{\odot}$ &  & ~\it Myr &  ~$s/am^{2}$  \\
\hline
4.5 & ~0.77 & ~8.9 & ~9.3 & ~685  & ~--1090 & ~--185 & ~230 & ~270 & ~6.1 & ~4.6 \\
3.5 & ~0.52 & ~10.8& ~9.9 & ~1276 & ~--1742 & ~--78  & ~410 & ~250 & ~6.3 & ~6.0 \\
5.0 & ~0.83 & ~9.1 & ~9.9 & ~1300 & ~--1492 & ~103   & ~240 & ~185 & ~9.8 & ~6.8 \\
\hline\\
\end{tabular}}
\end{table*}
%---------------------------------------------------------------------
\end{document}